\documentclass[a4paper,12pt]{article}

\usepackage{graphicx}

\begin{document}

\begin{center}
{\large\bf Transversal and longitudinal gluon spectral functions \\ 
from twisted mass lattice QCD with $N_f=2+1+1$ flavors} 

\vspace{0.4cm}
{E.-M. Ilgenfritz$^{1,\P}$}, {J.M. Pawlowski$^{2,3}$},{A. Rothkopf$^{4}$}, {A.M. Trunin$^{1}$}

\vspace{0.4cm}
$^1${BLTP, Joint Institute for Nuclear Research, Joliot-Curie str. 6, 141980 Dubna, Russian Federation}

$^2${ITP, Universit\"at Heidelberg, Philosophenweg 16, 69120 Heidelberg, Germany}

$^3${EMMI, GSI Helmholtz-Zentrum fuer Schwerionenforschung mbH, 64291 Darmstadt, Germany}

$^4${ITP, Universit\"at Heidelberg, Philosophenweg 12, 69120 Heidelberg, Germany}

$^\P${E-mail: ilgenfri@theor.jinr.ru}
\end{center}

\centerline{\bf Abstract}
I report on the first application of a novel, generalized 
Bayesian reconstruction (BR) method for spectral functions 
to the characterization of QCD constituents. These spectral functions find  
applications in off-shell kinetics of the quark-gluon plasma 
and in calculations of transport coefficients. The new BR method 
is applied to Euclidean propagator data, obtained in Landau 
gauge on lattices with $N_f=2+1+1$ dynamical flavors by the 
``twisted mass at finite temperature'' (tmfT) collaboration. 
The deployed reconstruction method is designed for 
spectral functions that can exhibit positivity violation (opposed to that of hadronic
bound states). The transversal and longitudinal gluon spectral 
functions show a robust structure composed of quasiparticle 
peak and a negative trough. Characteristic differences between 
the hadronic and the plasma phase and between the two channels 
become visible. We obtain the temperature dependence of the 
transversal and longitudinal gluon masses.
PACS: numbers.

\section{Introduction}

Nowadays, the focus of lattice studies of extreme QCD is increasingly shifting
from mere phase structure to an understanding of the real-time dynamics of 
strongly interacting matter. The aim is to learn how phase transitions
proceed and how the essentially different degrees of freedom emerge. 
Already since long quasiparticle ideas play a role in phenomenological 
descriptions of the quark-gluon plasma in equilibrium~\cite{ref1}, and later 
appeared in more ambitious off-shell kinetic approaches~\cite{ref2}. In these 
cases, the ``in-medium'' gluon and quark spectral functions are an important 
input. During the last years, functional approaches (FRG and DSE) to 
continuum QCD have been developed (for two recent FRG studies see ~\cite{ref3,ref3-1}). Among other objectives, this
work has pointed out how quasiparticle spectral functions can be used in order 
to express transport coefficients of the quark-gluon plasma~\cite{ref4,ref4a}.

In our work~\cite{ref5}, that I am going to report on, the spectral function 
encoding gluon properties is calculated in a way that is applicable across the phase 
transition (crossover) of lattice QCD. In our concrete case this method is
applied to simulation results of QCD with $N_f=2+1+1$ dynamical quarks 
pursued by the tmfT collaboration~\cite{ref6,ref6a,ref6b}.

The gluon spectral function witnesses the fact that the gluon is not a 
physical particle: the spectral function is (i) gauge dependent 
and it (ii) violates spectral positivity~\cite{ref7,ref7a}. 
The degree of violation is expected 
to be stronger in confinement than in the deconfinement phase. In the 
deconfined (quark-gluon plasma) phase, the quasiparticle picture of gluons 
and quarks is obviously useful. It comprises insight into the physics of 
the quark-gluon plasma complementary to what lattice gauge theory usually
provides. The violation of spectral positivity precludes the application 
of standard algorithms like MEM~\cite{ref8} or standard Bayesian 
reconstruction~\cite{ref9} for the calculation of the gluon spectral function.

\section{From Gluon Correlators to Gluon Spectral Functions}

Our raw data come from a (Euclidean) simulation of dynamical QCD within 
the twisted mass approach, including strange and charm quarks with physical
mass~\cite{ref6}. 
The simulations of the tmfT collaboration have been performed for
various light quark masses heavier than physical. In order to test our
reconstruction method we have selected ensembles generated for 
$m_{\pi} \approx 370~\mathrm{MeV}$ at three different lattice spacings
($\beta=1.90$, $\beta=1.95$ and $\beta=2.10$). 
I present here only the results for the finest lattice.
 
After gauge-fixing the lattice configurations to the Landau gauge, we have 
calculated the gluon propagator, extended to non-zero Matsubara frequencies. 
Unfolding the real-time information from this data is known to represent an 
ill-posed inverse problem. The results~\cite{ref5}, to be briefly discussed 
here, represent
the first physical case for which the novel formulation~\cite{ref10} of the 
Bayesian reconstruction (BR) procedure has been applied.
       
Primarily, the Euclidean gluon propagator (two-point function) is defined 
in terms of the 4d 
Fourier transform of the gauge field extracted from the links
\begin{equation}
A^a_\mu(x+\frac{\hat{\mu}}{2}) = \frac{1}{2iag_0} \left. \left(U_{x,\mu} - U^\dagger_{x,\mu}\right)\right|_{\mathrm{traceless}} \; ,
\end{equation} 
through the ensemble average 
\begin{equation}
D^{ab}_{\mu\nu}(q)=\langle \tilde{A}^a_\mu(q) \tilde{A}^b_\nu(-q)\rangle \; .
\end{equation} 
In a thermal state (heat bath) characterized by some temperature, the gluon
propagator can be decomposed into transversal (chromomagnetic) and 
longitudinal (chromoelectric) parts 
\begin{equation}
D^{ab}_{\mu\nu}(q_4,{\bf q})=\delta^{ab}\left(P^T_{\mu\nu}~D_T(q_4,{\bf q})
                              + P^L_{\mu\nu}~D_L(q_4,{\bf q}2)\right) \; .
\end{equation} 
with the projectors 
$P^L_{\mu\nu} = \left(\delta_{\mu\nu} - \frac{q_\mu q_\nu}{q^2} \right) 
- P^T_{\mu\nu}$ (longitudinal) and 
$P^T_{\mu\nu} = \left(1 - \delta_{\mu 4}\right) 
                 \left(1 - \delta_{\nu 4}\right) 
         \left( \delta_{\mu\nu} - \frac{q_\mu q_\nu}{{\bf q}^2} \right)$
(transversal).
These gluon correlators are related to the respective spectral functions via 
the K\"allen-Lehmann representation linking Minkowski and Euclidean space
\begin{eqnarray}
D_{T,L}(q_4,{\bf q}) &=& \int_{-\infty}^{\infty} \frac{1}{i q_4 + \omega}~
                         \rho_{T,L}(\omega,{\bf q}) d \omega \\ 
D_{T,L}(q_4,{\bf q}) &=& \int_{0}^{\infty} \frac{2 \omega}{q_4^2 + \omega^2}~  
                         \rho_{T,L}(\omega,{\bf q}) d \omega \; . 
\end{eqnarray} 
The spectral function is understood to be antisymmetric, 
$\rho(-\omega)=-\rho(\omega)$.

Inverting this relation for transversal and longitudinal propagators at
any ${\bf q}$ is an inverse problem, which can be tackled with Bayesian methods. 
In this approach, the probability functional of a test spectral function 
$\rho(\omega)$ can be written in factorized form 
\begin{equation}
P[\rho|D,I] \propto P[D|\rho,I] P[\rho|I] \; .
\end{equation}
The first factor $P[D|\rho,I]=\exp\left(- L\right)$ contains the $\chi^2$
likelihood $L$ that $\rho$ reproduces the lattice data $D_i$ (of $D_T$ or 
$D_L$) for each Matsubara frequency, 
\begin{equation}
L=\frac{1}{2} \sum_{i,j=1}^{N_{q_4}} (D_i - D^\rho_i) C_{ij}^{-1} 
                                     (D_j - D^\rho_j) \; ,
\end{equation}
where $C_{ij}$ is the covariance matrix of the measured $D_i$ and
\begin{equation}
D^\rho_i = \sum_{l=1}^{N_\omega} K_{il}~\rho_l~\Delta\omega_l ; \qquad 0 \leq i \leq N_{q_4}
\end{equation}
is the binned integral form of the K\"allan-Lehmann representation 
for $D_i$ (with a huge number of bins $N_\omega >> N_{q_4}$, compared to the 
small number of data points (Matsubara frequencies). This makes a 
regularization indispensable.

The second factor contains the so-called prior probability in the form
$P[\rho|I] = \exp\left(\alpha S\right)$. The prior data $I$ consists of
the functional form of $S=S[\rho(\omega),m(\omega)]$, which depends on 
the applied test spectral function $\rho$ and a default model, the 
model function $m(\omega)$, as well as the binning made to enable the 
integration. In the case of the Maximal Entropy Method (MEM) the 
Shannon-Jaynes (relative) entropy plays this role
\begin{equation}
S_{MEM} =  \int d \omega \left( \rho(\omega) - m(\omega) 
- \rho(\omega) \log \frac{\rho(\omega)}{m(\omega)} \right) \;   ,
\end{equation}
while in the case of standard Bayesian reconstruction
\begin{equation}
S_{BR} = \int d \omega \left(1-\frac{\rho(\omega)}{m(\omega)} 
+ \log \left[\frac{\rho(\omega)}{m(\omega)}\right]\right) \; .
\end{equation}
Both is not applicable if spectral positivity is violated. 
Instead we apply a generalized BR regulator~\cite{ref5}
\begin{equation}
S_{BR}^{g} = \int d\omega 
\left( -\frac{|\rho(\omega) - m(\omega)|}{h(\omega)}
+ \log \left[1 + \frac{|\rho(\omega) - m(\omega)|}{h(\omega)}\right] \right).
\end{equation}
The additional model function $h(\omega)$ presents the confidence in the
default model function $m(\omega)$.
The spectral reconstruction consists in finding the most probable 
spectral function according to
\begin{equation}
\left. \frac{\delta P[\rho|D,I]}{\delta \rho(\omega)}\right|_{\rho=\rho^{\mathrm{Bayes}}} 
= 0 \; .
\end{equation}
The generalized BR prior is the weakest among all similar priors, 
in contrast -- for example -- to a prior quadratic in 
$\rho(\omega) - m(\omega)$. 
Consequently, the spectral function is to a maximal amount determined by 
the lattice data $D_i$ and only minimally influenced by the regulator.
\begin{figure}
\begin{center}
\includegraphics[scale=0.30]{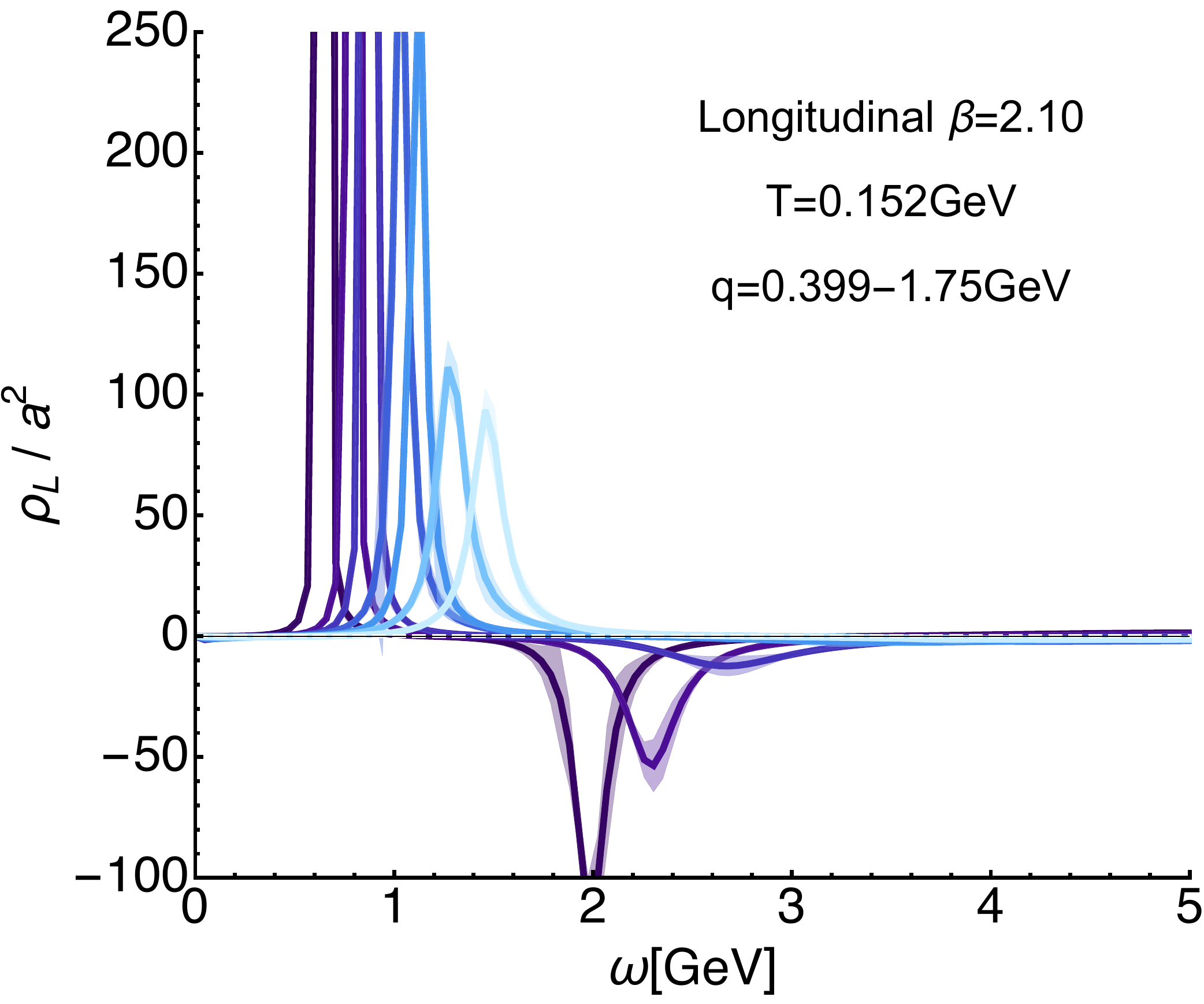}\hspace{0.5cm}
\includegraphics[scale=0.30]{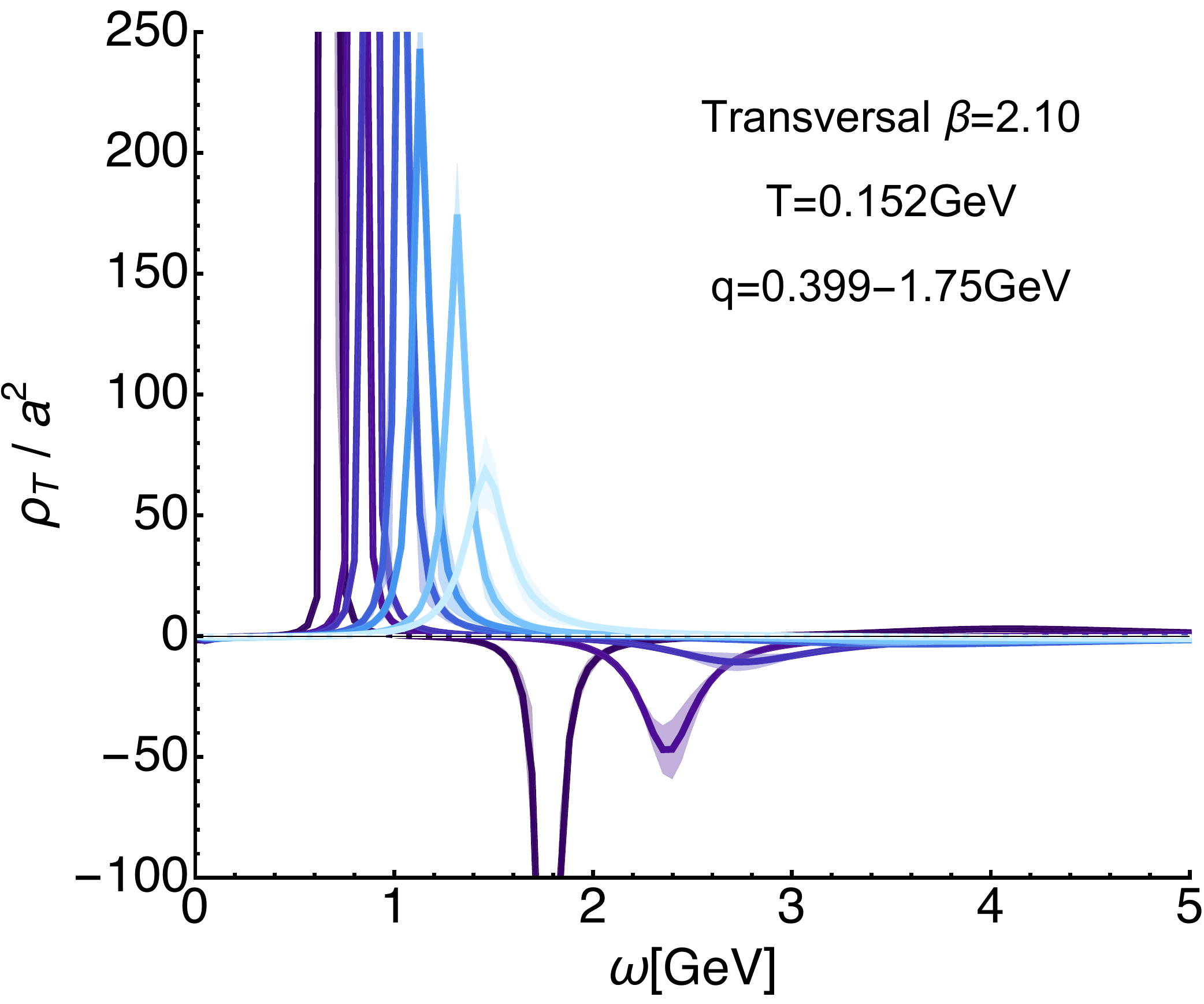}
\end{center}\vspace{-0.5cm}
\caption{The longitudinal and the transversal gluon spectral functions
for different momenta at temperature $T=152~\mathrm{MeV} < T_\chi$} 
\end{figure}
Previously, the $T$-dependence of the electric and magnetic gluon 
propagators~\cite{ref11,ref11a,ref11b} 
was discussed with attention restricted only to the gluon propagators at 
Matsubara frequency $q_4=0$. 
Here we are going to exploit the splitting of the 
propagator data between Matsubara frequencies. From this we shall infer
the corresponding $\omega$-dependence of the reconstructed spectral functions. 
In previous attempts to the reconstruction problem it has been often assumed 
that the propagator satisfies some $O(4)$ invariance~\cite{ref12} 
in momentum space,
$D_{T,L}\left(q_4,|{\bf q}|\right) \approx D_{T,L}\left(0,\sqrt{q_4^2 + {\bf q}^2} \right)$.
As our correlator has shown, this assumption is justified within small deviations only close 
to $q_4 = 0$, but becomes problematic near to the end of the Brillouin zone.

\section{Reconstructed spectra below and above $T_\chi$}

I show in Fig. 1 the result of reconstructing the longitudinal and 
transversal gluon spectral functions for $T=0.152~\mathrm{GeV} < T_\chi$
from data obtained on a lattice with $N_\sigma=48$ and a temporal extent 
$N_\tau=20$ at $m_\pi=370~\mathrm{MeV}$ and $\beta=2.10$:
We clearly observe at low temperature a structure of a peak and subsequent
broader negative trough in both channels. The negative contribution appears 
slightly stronger in the transversal sector.

Next I show in Fig. 2 the result of reconstructing the longitudinal and 
transversal gluon spectral functions for $T=0.305~\mathrm{GeV} > T_\chi$
from a similar data set with $N_\sigma=32$ and $N_\tau=10$: 
\begin{figure}
\begin{center}
\includegraphics[scale=0.30]{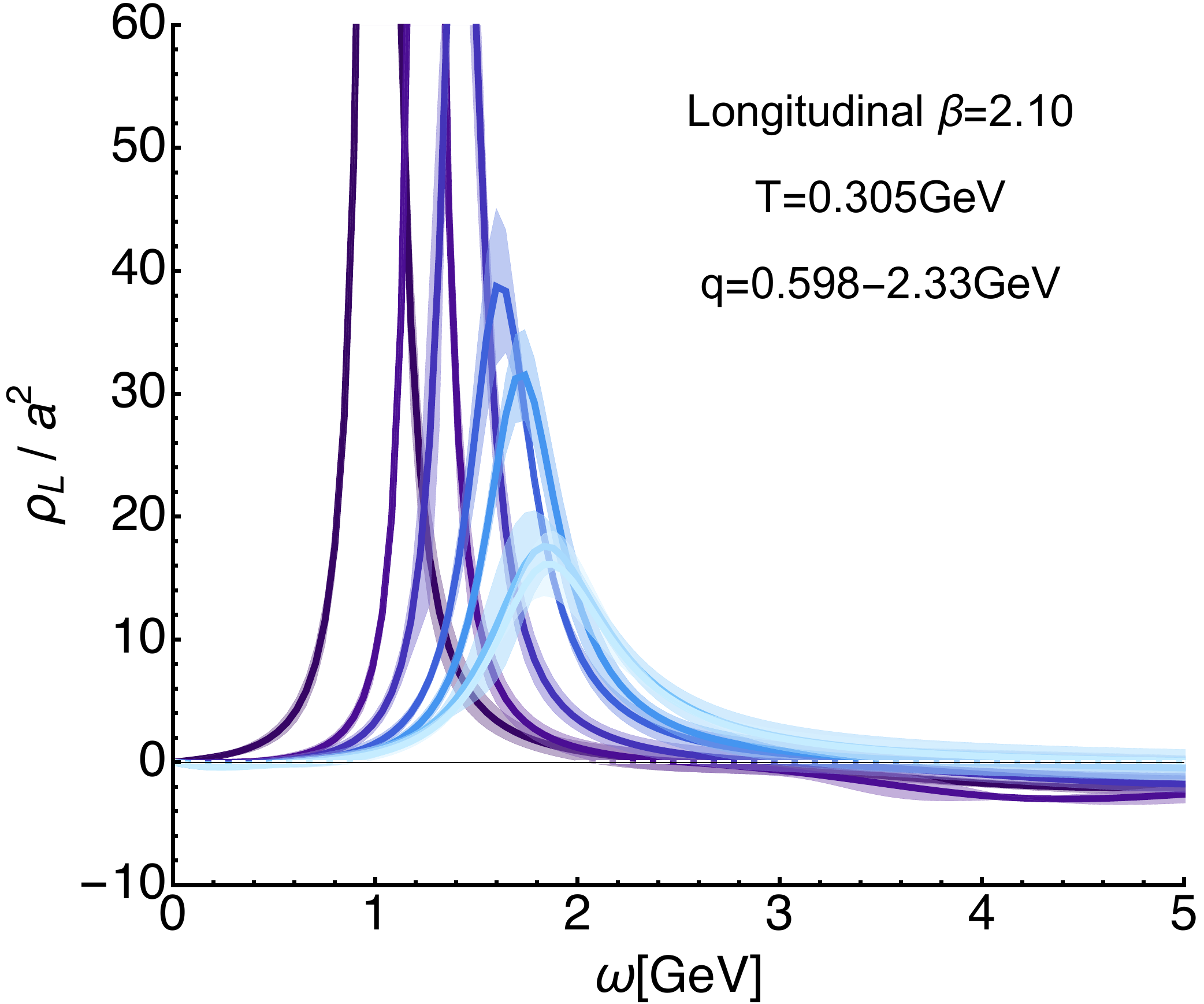}\hspace{0.5cm}
\includegraphics[scale=0.30]{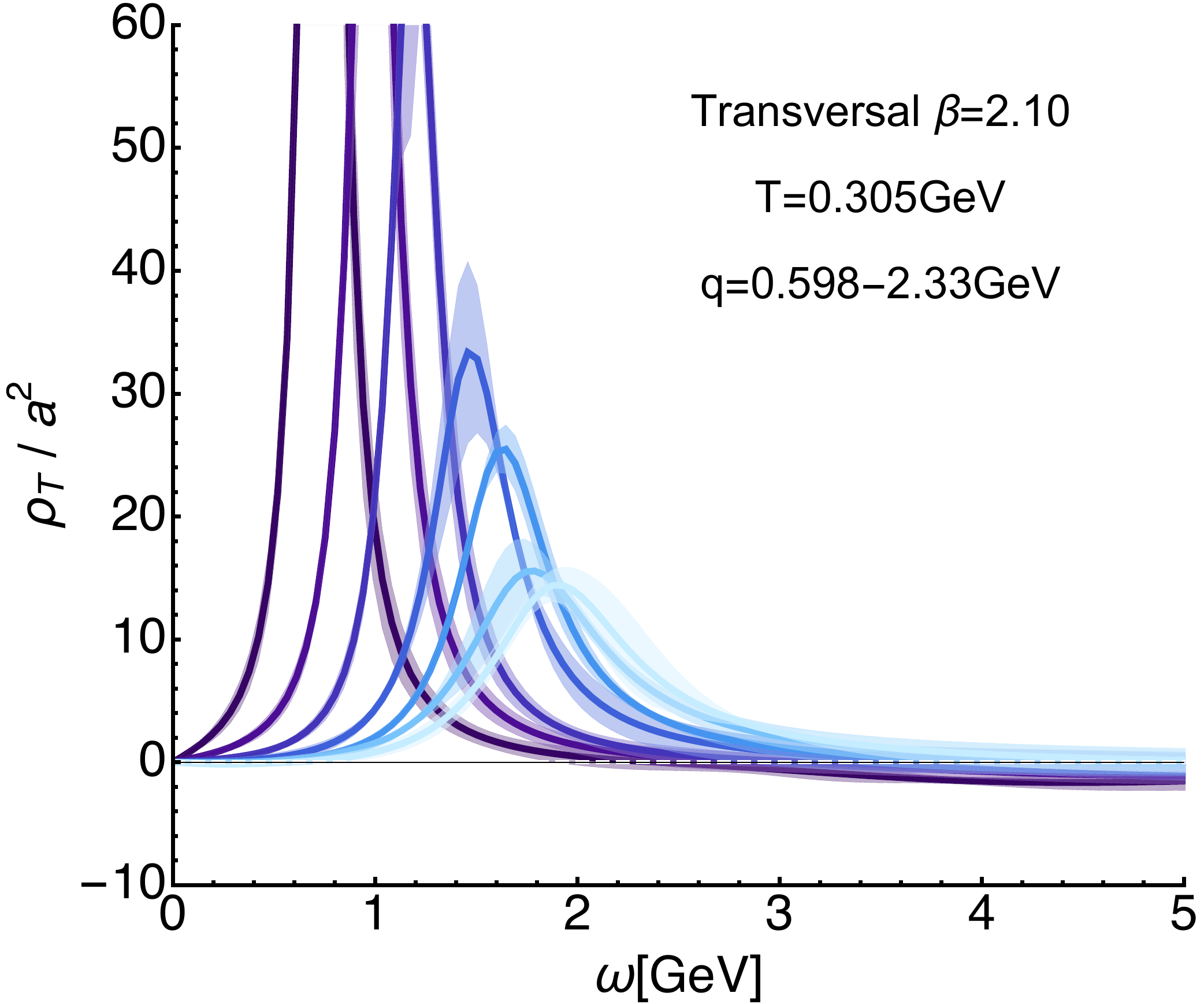}
\end{center}\vspace{-0.5cm}
\caption{The longitudinal and the transversal gluon spectral functions
for different momenta at temperature $T=305~\mathrm{MeV} > T_\chi$}
\end{figure}
The negative contribution is significantly reduced at this temperature 
$T > T_\chi$ in both channels.

One may use the peak position of the low-momentum structure of the longitudinal
spectral function to define a longitudinal gluon dispersion relation 
$\omega^{\rm max}_{L}({\bf q})$ valid at the respective temperature. 
In Fig. 3 (left) I show the maximum position in energy vs. momentum for 
eight values of the temperature. Fig. 3 (right) shows a free--field fit for 
the lowest and the highest temperature. The intercept at $|{\bf q}|=0$ 
(the longitudinal gluon mass) is decreasing with rising temperature across 
the transition (which is, actually, a thermal crossover). Even for the highest 
temperature, the longitudinal gluon mass is larger than the Debye screening
mass exhibited by the heavy quark potential at the same temperature in
simulations with $N_f=2+1$ flavors~\cite{ref13} (with no momentum 
dependence). 
A corresponding measurement of the Debye mass is not yet available from 
our twisted mass simulations with $N_f=2+1+1$.

In Fig. 4 the transversal gluon dispersion relation 
$\omega^{\rm max}_{T}({\bf q})$ is shown. 
In the left panel I show the maximum position in energy vs. momentum for 
eight temperatures. Fig. 4 (right) shows a free--field fit for the lowest 
and the highest temperature.  
\begin{figure}[h]
\begin{center}
\includegraphics[scale=0.50, trim= 0 0.5cm 0 0, clip=true]{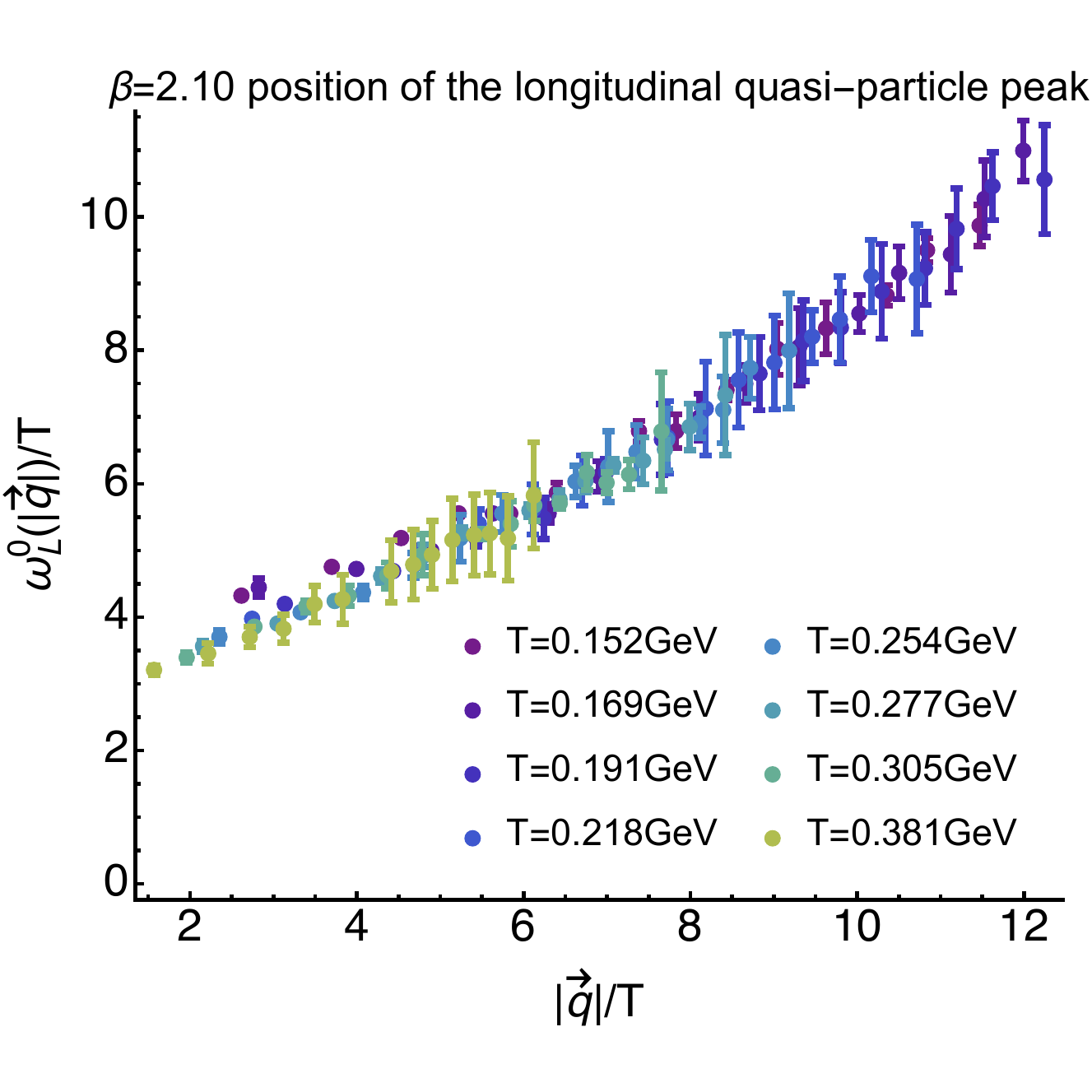}\hspace{0.5cm}
\includegraphics[scale=0.50]{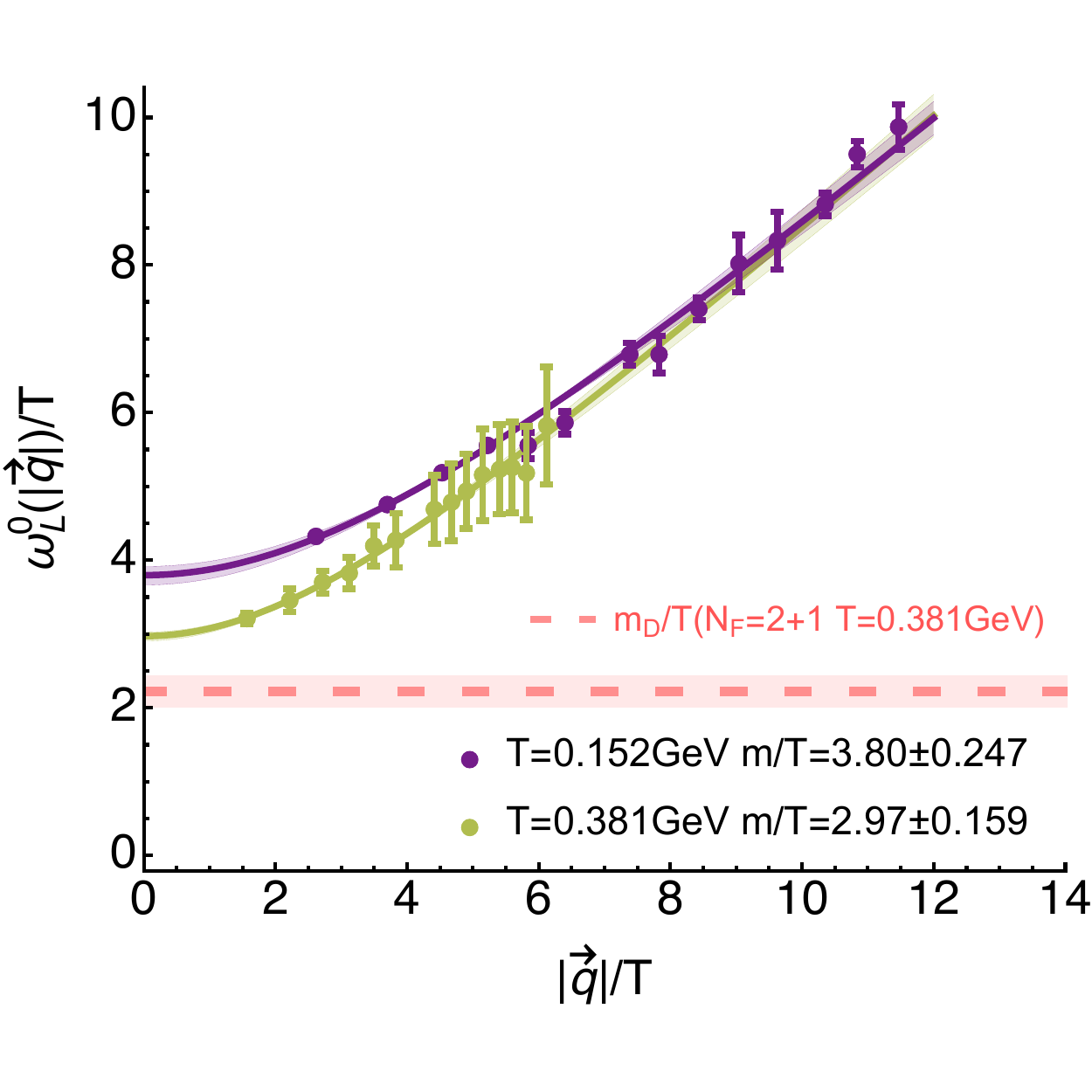}
\end{center}\vspace{-0.5cm}
\caption{Left: Momentum dependence of the longitudinal quasiparticle
peak position at $\beta=2.10$ showing a non-zero intercept.
Right: Fit of the lowest and highest temperature curves with the 
free--field ansatz $\omega_L^0(|\vec{q}|)=A\sqrt{B^2+|\vec{q}|^2}$. 
The gluon quasiparticle mass is defined as $m=AB$.
The Debye mass from $N_f=2+1$ lattice QCD~\cite{ref13} is given for 
comparison.}
\end{figure}

\begin{figure}
\begin{center}
\includegraphics[scale=0.50, trim= 0 0.5cm 0 0, clip=true]{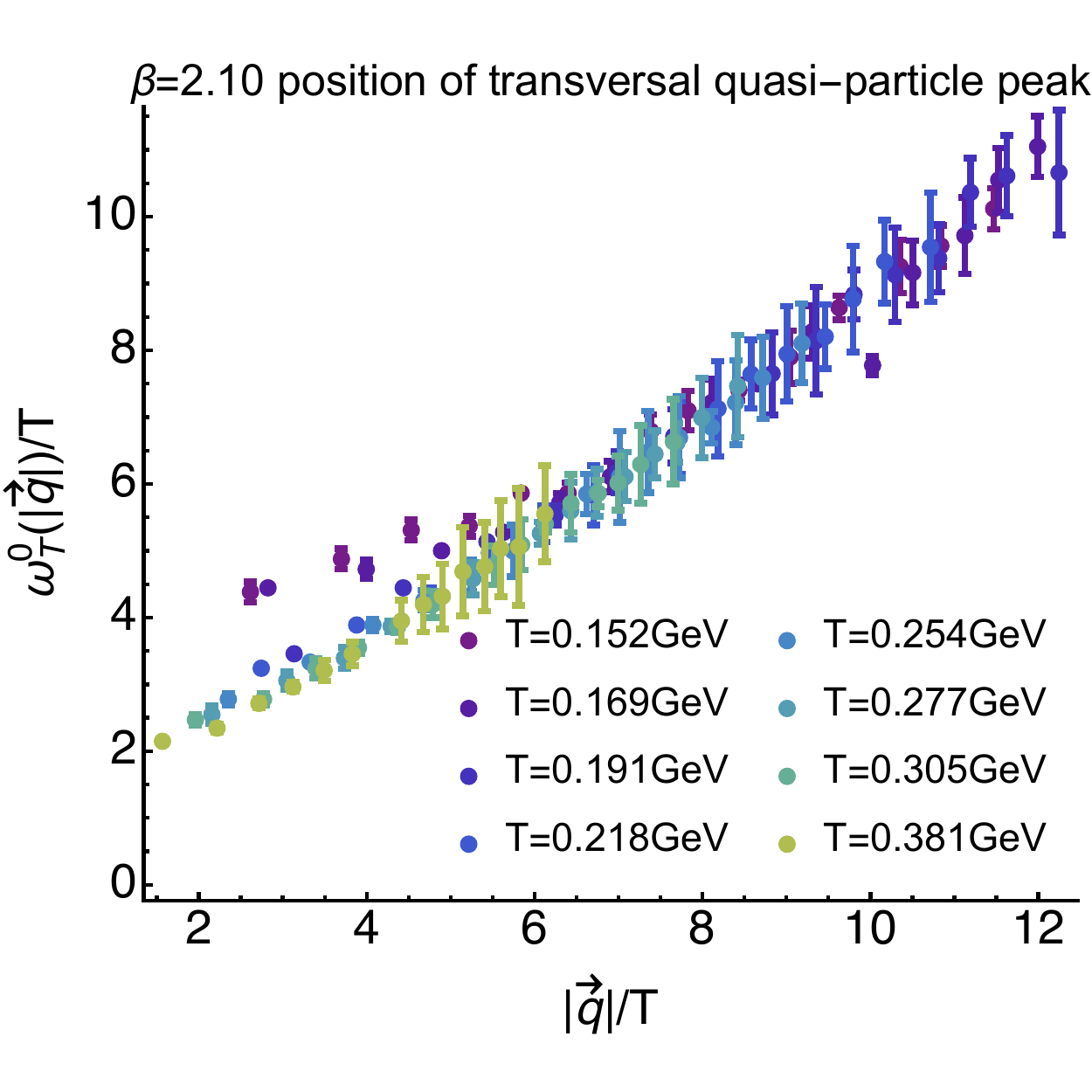}\hspace{0.5cm}
\includegraphics[scale=0.50]{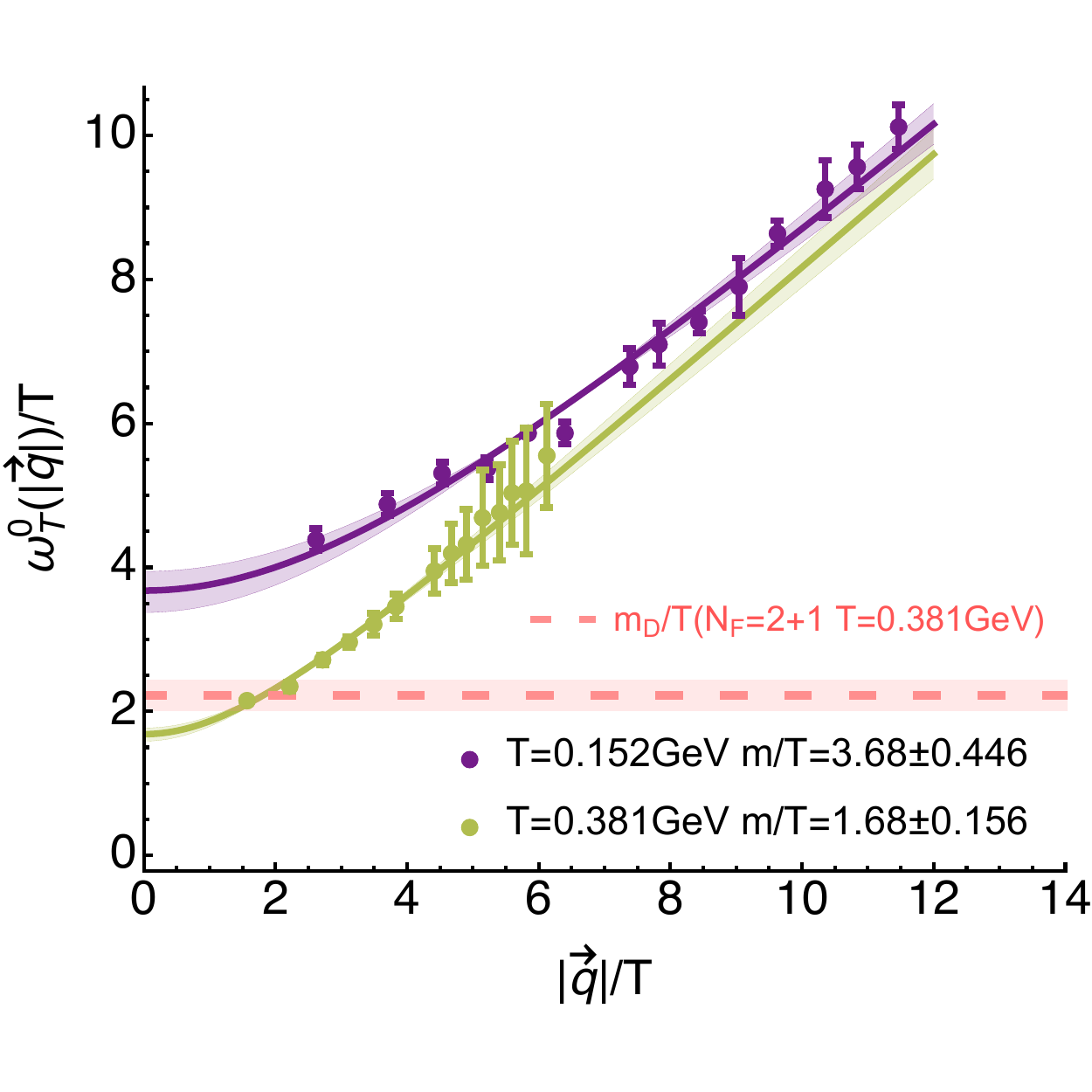}
\end{center}\vspace{-0.5cm}
\caption{Left: Momentum dependence of the transversal quasiparticle
peak position at $\beta=2.10$ showing a non-zero intercept.
Right: Fit of the lowest and highest temperature curves with the 
free--field ansatz $\omega_L^0(|\vec{q}|)=A\sqrt{B^2+|\vec{q}|^2}$. 
The gluon quasiparticle mass is defined as $m=AB$.
The Debye mass from $N_f=2+1$ lattice QCD~\cite{ref13} is given for 
comparison.}
\end{figure}

Also in the transversal case, the intercept at $|{\bf q}|=0$ 
(the transversal gluon mass) is decreasing with rising temperature across 
the transition. Now, for the higher temperature, the transversal gluon 
mass drops below the Debye screening mass of the heavy quark potential 
at the same temperature obtained in simulations with $N_f=2+1$ 
flavors~\cite{ref13}.

\section{Conclusion and Outlook}

We have demonstrated the successful application of a new Bayesian 
method to reconstruct the gluon spectral function from gluon two-point 
correlator data in $N_f=2+1+1$ flavor twisted mass QCD thermodynamics
with $m_{\pi} = 370~\mathrm{MeV}$ across the deconfining and chiral 
restoration crossover.

Presently, we are preparing to repeat this analysis for twisted mass 
ensembles with a lower pion mass $m_{\pi} \approx 210~\mathrm{MeV}$.
A measurement of the static $Q\bar{Q}$ potential with $N_f=2+1+1$
twisted mass flavors in the deconfinement phase and the determination of the 
Debye screening mass belongs to our nearest goals in a program of twisted 
mass lattice QCD thermodynamics.

\section*{Acknowledgment}

We are grateful to the tmfT collaboration for the lattice configuration and
to the HybriLIT team of JINR's Laboratory for Information Technology for 
the opportunity to perform the gauge fixing at their hybrid cluster. This work is 
part of and supported by the DFG Collaborative Research Centre "SFB 1225 (ISOQUANT)".


\begin{thebibliography}{99}

\bibitem{ref1}
{\it Cassing W., Bratkovskaya E.L.}
Parton transport and hadronization from the dynamical quasiparticle point 
of view //
Phys. Rev. C78 (2008) 034919, arXiv:0808.0022 [hep-ph]

\bibitem{ref2}
{\it Berrehrah H., Bratkovskaya E., Steinert T., Cassing W.}
A dynamical quasiparticle approach for the QGP bulk and transport properties //
Int. J. Mod. Phys. E25 (2016) no.07, 1642003, arXiv:1605.02371 [hep-ph]

\bibitem{ref3}
{\it Cyrol A. K., Mitter M., Pawlowski J. M., Strodthoff N.}
Non-perturbative finite-temperature Yang-Mills theory //
arXiv:1708.03482 [hep-ph].

\bibitem{ref3-1}
{\it Cyrol A. K., Mitter M., Pawlowski J. M., Strodthoff N.}
Non-perturbative quark, gluon and meson correlators of unquenched QCD //
arXiv:1706.06326 [hep-ph].

\bibitem{ref4}
{\it Christiansen N., Haas M., Pawlowski J.M., Strodthoff N.}
Transport Coefficients in Yang--Mills Theory and QCD //
Phys. Rev. Lett. 115 (2015) 112002, arXiv:1411.7986 [hep-ph] 

\bibitem{ref4a}
{\it Haas M., Fister L., Pawlowski J.M.}
Gluon spectral functions and transport coefficients in Yang--Mills theory //
Phys. Rev. D90 (2014) 091501, arXiv:1308.4960 [hep-ph]

\bibitem{ref5}
{\it Ilgenfritz E.-M., Pawlowski J.M., Rothkopf A., Trunin A.}
Finite temperature gluon spectral functions from $N_f=2+1+1$ lattice QCD //
arXiv:1701.08610 [hep-lat]

\bibitem{ref6}
{\it Burger F., Hotzel G., M\"uller-Preussker M., Ilgenfritz E.-M., 
Lombardo M.P.}
Towards thermodynamics with Nf=2+1+1 twisted mass quarks //
PoS Lattice2013 (2013) 153, arXiv:1311.1631 [hep-lat]  

\bibitem{ref6a}
{\it Burger F., Ilgenfritz E.-M., Lombardo M.P., M\"uller-Preussker M., Trunin A.} 	
Towards the quark--gluon plasma Equation of State with dynamical strange and 
charm quarks //
J. Phys. Conf. Ser. 668 (2016) no.1, 012092, arXiv:1510.02262 [hep-lat]  

\bibitem{ref6b} 	
{\it Trunin A., Burger F., Ilgenfritz E.-M., Lombardo M.P., M\"uller-Preussker M.} 	
Topological susceptibility from $N_f=2+1+1$ lattice QCD at nonzero temperature //
J. Phys. Conf. Ser. 668 (2016) no.1, 012123, arXiv:1510.02265 [hep-lat]  

\bibitem{ref7}
{\it Alkofer R., von Smekal L.}
The Infrared behavior of QCD Green's functions: Confinement dynamical 
symmetry breaking, and hadrons as relativistic bound states //
Phys. Rept. 353 (2001) 281, hep-ph/0007355  

\bibitem{ref7a}
{\it Cornwall J.M.}
Positivity violations in QCD //
Mod. Phys. Lett. A28 (2013) 1330035, arXiv:1310.7897 [hep-ph]  

\bibitem{ref8}
{\it Asakawa M., Hatsuda T., Nakahara Y.}
Maximum entropy analysis of the spectral functions in lattice QCD //
Prog. Part. Nucl. Phys. 46 (2001) 459, hep-lat/0011040  

\bibitem{ref9}
{\it Burnier Y., Rothkopf A.}
Bayesian Approach to Spectral Function Reconstruction for Euclidean Quantum 
Field Theories //
Phys. Rev. Lett. 111 (2013) 182003, arXiv:1307.6106 [hep-lat]  

\bibitem{ref10}
{\it Rothkopf A.}
Bayesian inference of nonpositive spectral functions in quantum field theory //
Phys. Rev. D95 (2017) 056016, arXiv:1611.00482 [hep-ph] 

\bibitem{ref12}
{\it Dudal D., Oliveira O., Silva P.J.}
K\"allen-Lehmann spectroscopy for (un)physical degrees of freedom //
Phys. Rev. D89 (2014) 014010 arXiv:1310.4069 [hep-lat]

\bibitem{ref11}
{\it Aouane R., Bornyakov V.G., Ilgenfritz E.-M., Mitrjushkin V.K., 
M\"uller-Preussker M., Sternbeck A.}
Landau gauge gluon and ghost propagators at finite temperature from quenched 
lattice QCD //
Phys. Rev. D85 (2012) 034501, arXiv:1108.1735 [hep-lat]  

\bibitem{ref11a}
{\it Aouane R., Burger F., Ilgenfritz E.-M., M\"uller-Preussker M., 
Sternbeck A.}
Landau gauge gluon and ghost propagators from lattice QCD with $N_f=2$
twisted mass fermions at finite temperature //
Phys. Rev. D87 (2013) 114502, arXiv:1212.1102 [hep-lat] 

\bibitem{ref11b}
{\it Fischer, C.S., Fister L., Luecker J., Pawlowski J.M.}
Polyakov loop potential at finite density //
Phys. Lett. B732 (2014) 273, arXiv:1306.6022 [hep-ph]  

\bibitem{ref13}
{\it Burnier Y, Kaczmarek O., Rothkopf A.}
Quarkonium at finite temperature: Towards realistic phenomenology from first principles //
JHEP 1512 (2015) 101, arXiv:1509.07366 [hep-ph].

\end{thebibliography}
\end{document}